\documentclass[aps,pra,reprint,showpacs,amssymb,amsmath]{revtex4-1}
\pdfoutput=1
\usepackage{graphicx}
\usepackage{times}
\usepackage{color}
\usepackage{siunitx}
\usepackage{braket}
\newcommand{\Rb}{$^{87}$Rb }

\begin{document}
\title{Single- and two-qubit quantum gates using superimposed optical lattice potentials}
\author{Nils B. J\o rgensen}
\author{Mark G. Bason}
\author{Jacob F. Sherson}\email[]{sherson@phys.au.dk}
\affiliation{Department of Physics and Astronomy, Aarhus University, DK-8000 Aarhus C, Denmark}

\date{\today}

\begin{abstract}
Steps towards implementing a collision based two-qubit gate in optical lattices have previously been realized by the parallel merging all pairs of atoms in a periodicity two superlattice. In contrast, we propose an architecture which allows for the merger of a selected qubit pair in a novel long-periodicity superlattice structure consisting of two optical lattices with close-lying periodicity. We numerically optimize the gate time and fidelity, including the effects on neighboring atoms, and in the presence of experimental sources of error. Furthermore, the superlattice architecture induces a differential hyperfine shift, allowing for single-qubit gates. The fastest possible single-qubit gate times, given a maximal tolerable rotation error on the remaining atoms at various values of the lattice wavelengths, are identified. We find that robust single- and two-qubit gates with gate times of a few 100~$\mu$s and with error probabilities $\sim{}10^{-3}$ are possible.
\end{abstract}
\pagestyle{plain}
\maketitle

\section{Introduction}
The ability to prepare and manipulate ultracold atoms in optical lattices has led to many breakthroughs in the last decade. From demonstrating the superfluid to Mott-insulator transition~\cite{Greiner2002}, to strongly interacting Fermi gases~\cite{Chin2004,Zwierlein2005}, the purity and controllability of ultracold atoms has greatly benefited many-body physics~\cite{Bloch2012}. Due to the inherent, repeating pattern of an optical lattice and the long-coherence times of neutral atoms arranged in such systems, they are also viable candidates for quantum computing \cite{Brennen1999,Jaksch1999}. Ultracold atoms in optical lattices are scalable and offer parallelism due to their geometry~\cite{Nelson2007}. Implementing the two-qubit gates necessary for quantum computation is a long-standing problem using this approach. In optical lattices, two-qubit gates have been proposed~\cite{Charron2002} and conducted on many pairs of atoms in parallel~\cite{Anderlini2007,Mandel2003}, by making use of ground state collisions~\cite{Jaksch1999}.
Alternatively, one may make use of dipole-dipole interactions between Rydberg states \cite{Jaksch2000a,Muller2009,Møller2008}, as indicated by recent experiments on pairs of atoms in dipole traps \cite{Isenhower2010,Wilk2010}, or by means of hybrid atom-molecule schemes in optical lattices~\cite{Kuznetsova2010}.
 \begin{figure}[!ht]
		\includegraphics[width=\columnwidth]{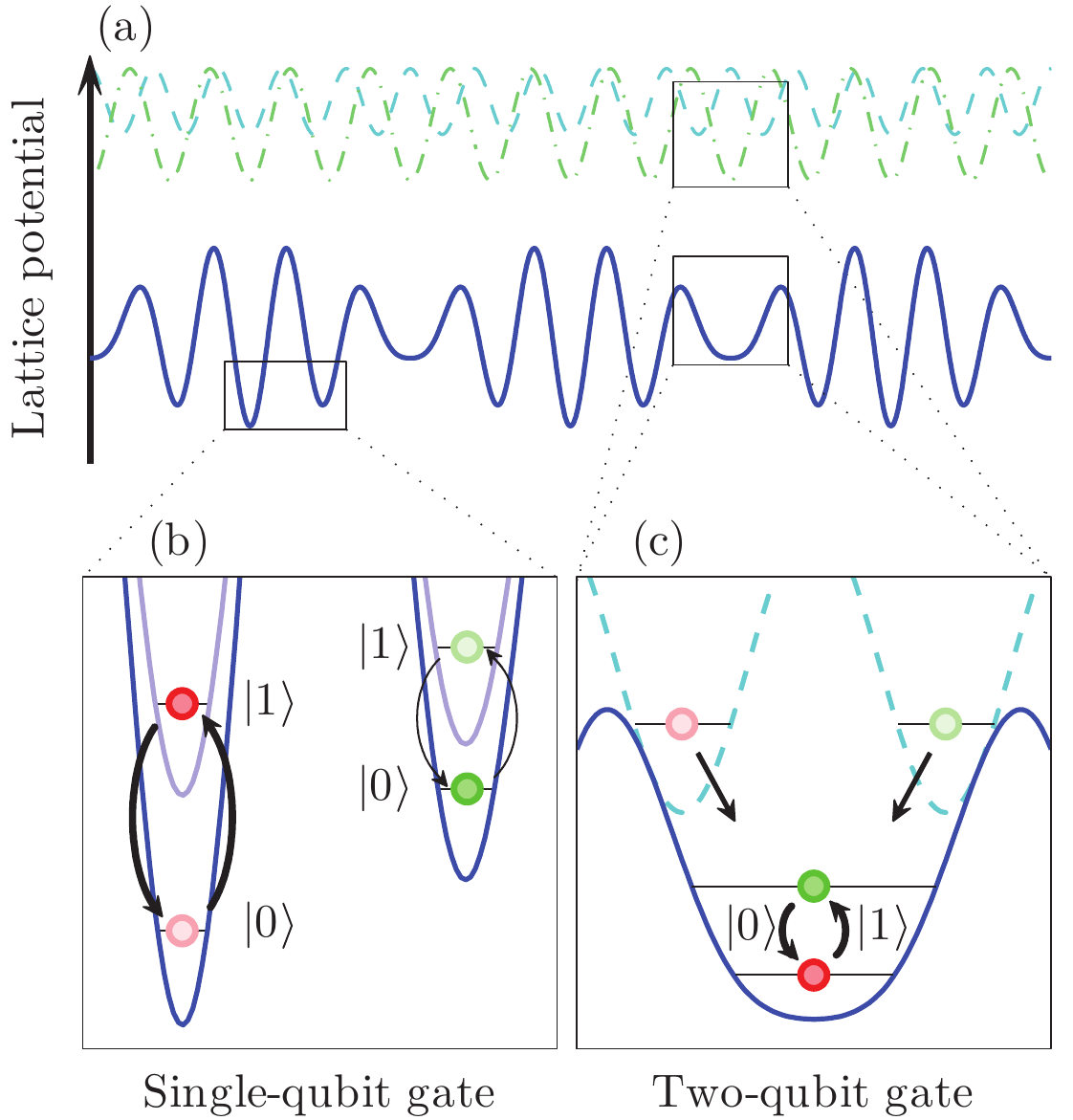}
	\caption{(Color online) Overview of the superlattice and the single- and two-qubit spin-state gates. (a) Two optical lattice potentials are superimposed to to create a long-period superlattice. (b) The varying well depth throughout the superlattice results in a varying spin transition frequency for each atom. A microwave tuned to the transition of one atom (red), only partially switches another atom (green). If the partially switched population is kept sufficiently low, a single-qubit gate is realized. (c) When a lattice potential of longer wavelength is added to a lattice potential of shorter wavelength, two wells, each holding an atom, can be merged. Through control of phase and well depth, atoms are sent into the vibrational ground and first excited state, where they interact for an arbitrary amount of time, before reversing the process. The interaction causes a spin state exchange resulting in a two-qubit gate.}
	\label{fig:overview}
\end{figure}
The challenge of implementing a two-qubit gate on a selected pair of atoms in a large array has yet to be fulfilled largely due to the experimental difficulty in obtaining an imaging resolution comparable to the lattice spacing. Initially this lead to proposals to achieve single site addressing using sub-diffraction optical techniques~\cite{Saffman2004,Cho2007,Gorshkov2008} and an experimental demonstration using   magnetic gradients~\cite{Schrader2004}. Recently, however, single site imaging~\cite{Bakr2009,Sherson2010} and single site addressing using a strongly focussed optical tweezer~\cite{Weitenberg:2011} were achieved. This paves the way for the realization of the two-qubit gates proposed for controllable micropotentials~\cite{Calarco2000,Briegel2000,Negretti2005,Dorner2005,Cirone2005,Charron2006,Negretti2012}, triple-wells~\cite{Tichy2013,Gajdacz2012}, and in optical lattices~\cite{You2000,Calarco2004,Weitenberg2011b}. As demonstrated in Ref.~\cite{Weitenberg2011b}, a high fidelity realization of gates using an optical tweezer imposes rather strict demands on the pointing stability of the addressing laser. One solution to the problem may involve the use of superlattices involving two optical lattices of separate optical frequencies. Such an arrangement has already been used to investigate double-well dynamics~\cite{Anderlini2007,Folling2007a} and demonstrate patterned loading~\cite{Peil2003a} in a triple-well superlattice. Recently, an additional long-period addressing lattice, superimposed on a conventional short-period lattice with an atomic filling of roughly one percent, has allowed the demonstration of unitary single-qubit gates with a fidelity of around 96\%~\cite{Lee2013}. Similar techniques have also been used in optical cavities with multiple wavelength lattices~\cite{Brahms2011,Botter2013}.

We propose a superlattice architecture in which both single- and two-qubit gates can be performed. The large period nature of our superimposed potentials allows selective addressing of individual lattice sites periodically spaced throughout the lattice as seen in Fig.~\ref{fig:overview}(b) in analogy with the architecture presented in Ref.~\cite{Lee2013}. Here we present detailed calculations of the achievable gate speed versus lattice frequency and in particular focus on the optimum trade-off between speed and the detrimental effect of spontaneous emission.
Two-qubit gates are facilitated by the merger and interaction of two initially separated atoms via spin-exchange as seen in Fig.~\ref{fig:overview}(c). For both single- and two-qubit gates we demonstrate errors below $10^{-3}$ including experimental sources of error.

This paper is organized along the following lines. The combination of two optical lattice potentials to form a superlattice is introduced in section~\ref{sec:potential}. The ability to perform single-qubit gates by exploiting the differential ac-Stark shift is discussed in section~\ref{sec:single}. The two-qubit gate using local collisional interactions is the subject of section~\ref{sec:two}, where numerical optimization is applied to determine minimum gate time and maximal fidelity. Section~\ref{sec:conc} summarises the paper's conclusions and highlights perspectives for the future.

\section{Long period superlattice potential}
\label{sec:potential}
The dipole potential experienced by a ground-state alkali atom in an optical field with wavelength $\lambda$ is~\cite{Grimm},
\begin{equation}
\begin{split}
& U(I(\textbf{r}),\lambda,\mathcal{P})
\\ = & \frac{\pi c^2 \Gamma}{2 \omega_0^3} \left( \frac{2+\mathcal{P} g_F m_F}{\Delta_{2,F} (\lambda) } + \frac{1-\mathcal{P} g_F m_F}{\Delta_{1,F}(\lambda)} \right) I(\textbf{r}).
\label{eq:basic_potential}
\end{split}
\end{equation}
Here the optical polarization $\mathcal{P}=0, \pm 1$  for linearly and circularly $\sigma^\pm$ polarized light respectively, $g_F$ is the Land\'{e} factor and $m_F$ the magnetic quantum number. $\Delta_{i,F} (\lambda)$ is the laser detuning given by $\Delta_{i,F}(\lambda) = \omega_\text{laser}(\lambda) - \omega_{i,F}$ where $i = 1,2$ refers to the $\text{D}_1$ and $\text{D}_2$ lines. This equation is valid for large detunings such that $\Delta_{i,F} (\lambda)>\Delta_\text{HFS}$, the excited-state hyperfine splitting. In the case of two counter-propagating fields an optical lattice with a lattice spacing $a_\text{lat} = \lambda/2$ is formed.

Adding two optical lattice potentials of similar wavelength light creates a 1D long-period superlattice with potential wells of varying depth, as seen in Fig.~\ref{fig:overview}(a). The length of one superlattice period (SLP) is $a_\text{SLP} = (\lambda_2^{-1} - \lambda_1^{-1})^{-1}/2$, three SLPs are seen in Fig.~\ref{fig:overview}(a). In this work, we consider a SLP in which the longer period lattice passes through one less cycle than the shorter period lattice, leading to the relation $\lambda_1/\lambda_2 = (n-1)/n$, where $n$ is the number of cycles in a SLP with $\lambda_2 < \lambda_1$.

\section{Single-qubit gate}\label{sec:single}

Throughout this work, we treat an array of single \Rb atoms confined to lattice sites as our starting point. Such a situation is readily realized through use of the superfluid to Mott-insulator transistion~\cite{Greiner2002}. The different spin states $\vert 0 \rangle \equiv \vert F=1,1 \rangle$ and $\vert 1 \rangle \equiv \vert F=2,2 \rangle$ experience different potentials when using $\sigma^\pm$ polarized light, as shown in Eq. (\ref{eq:basic_potential}). The varying intensity of each well in a SLP, causes the hyperfine transition $\Delta U_i = U_i(\Ket{1}) - U_i(\Ket{0})$ to differ for different atoms in lattice sites $i$. If a microwave $\pi$--pulse tuned to switch a target atom $j$ is applied throughout the superlattice, the population $P_i$ of all the atoms will oscillate $P_i = \tfrac{1}{2}  \left(\tfrac{\chi_i}{\Omega_i}\right)^2 [1 - \cos^2 (\Omega_i t)]$, where $\chi_i$ is the Rabi frequency, the generalized Rabi frequency $\Omega_i = \big ( \chi_i^2 + {\Delta^{ij}_\text{R}}^2 \big ) ^{1/2}$ and $\Delta^{ij}_\text{R}= (\Delta U_i - \Delta U_j)/\hbar$ is the detuning of the transition of atom $i$ compared to the transition of target atom $j$. Atoms in the selected wells are switched through a $\pi$--pulse, while each of the  other atoms of the SLP are kept beneath a threshold population $P_\text{t} = (\chi_k/\Omega_k)^2$ where $k$ denotes the site with minimal detuning. The detuning can be expressed through the threshold population $\vert \Delta_\text{R}^{ij} \vert = \chi_i [(1 - P_\text{t})/P_\text{t}]^{1/2} \approx \chi_i P_\text{t}^{-1/2}$, with the approximation being valid for $P_\text{t} \ll 1$. For a given threshold population, the $\pi$--pulse duration used to address the target atom can then be calculated $t_\text{a} = \pi /\sqrt{P_\text{t}}\vert \Delta_\text{R}^{ij} \vert$.
Finding the fastest possible gate time thus reduces to calculating the detunings $\Delta_\text{R}^{ij}$ for all atoms $i \neq j$ in a SLP. We note that this is of course a conservative approach: with the detailed know\-ledge of all detunings in a SLP one may also engineer pulse durations that produce less residual excitation than $P_\text{t}$.

The one dimensional potential for atoms in the field of the two standing waves comes through Eq. (\ref{eq:basic_potential}) . The primary laser potential depth is one unit of recoil energy $E_\text{r} (\lambda) = h^2/2m\lambda^2$, while the wavelength and relative intensity of the secondary laser is varied through a scaling parameter $A$. The total potential is thus given by
\begin{equation}
\begin{split}
& \frac{U_\text{SL}(x,\eta,A,\lambda_2,\mathcal{P}_1,\mathcal{P}_2)}{E_\text{r} (\lambda_1)}
\\ = & - \eta \left[ \cos^2(k_1 x) + A \frac{U(\lambda_2,\mathcal{P}_2)}{U(\lambda_1,\mathcal{P}_1)} \cos^2(k_2 x) \right] ,
\label{eq:potential_num}
\end{split}
\end{equation}
where $k_{1,2}$ are the wave numbers of the two lattice beams and $\eta$ is an additional scaling factor. The minus sign arises from the fact that only red detuned light is taken into consideration.

\begin{figure}[tb]
		\includegraphics[width=\columnwidth]{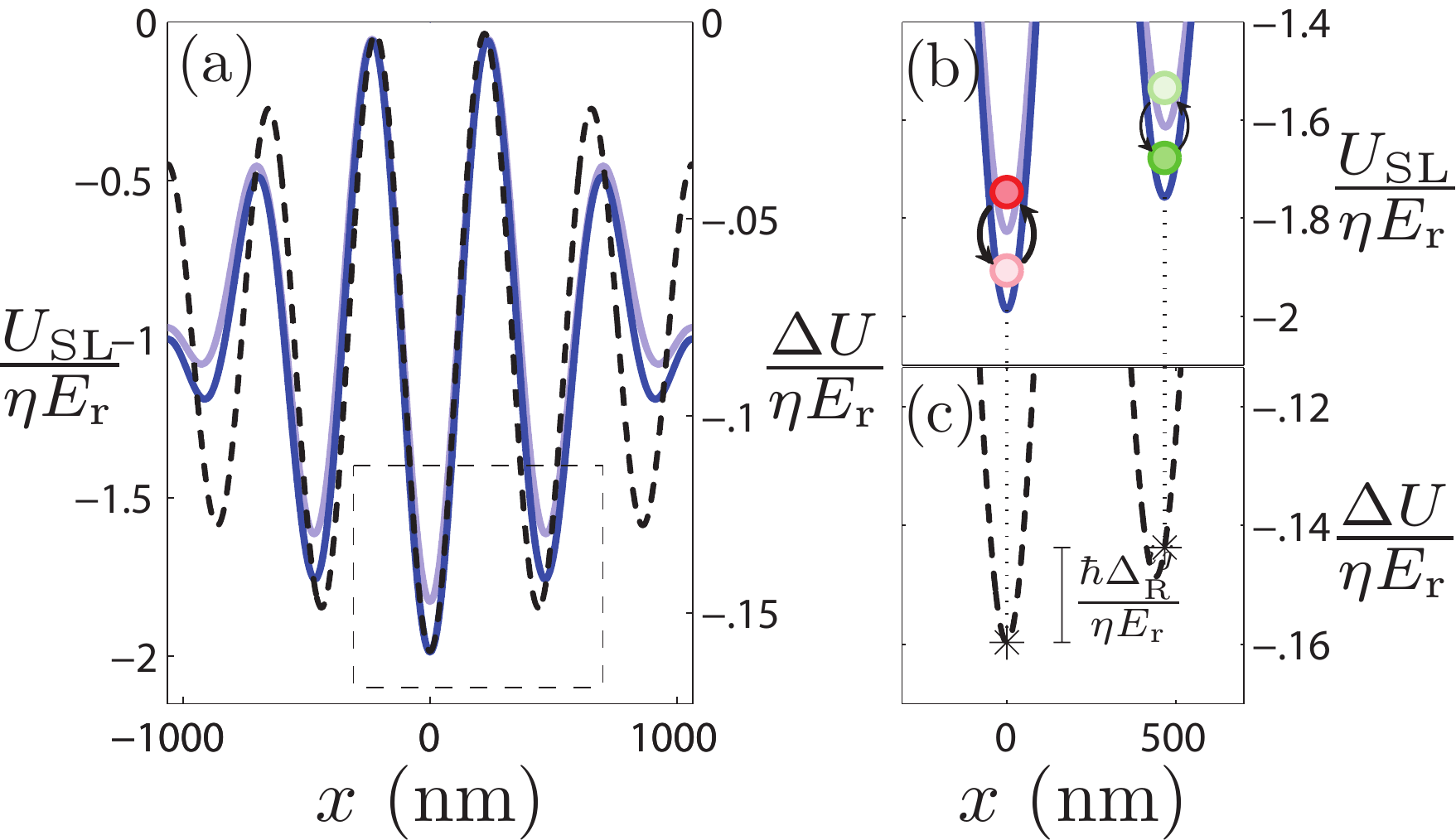}
	\caption{(Color online) A schematic overview of the mechanism of the single-qubit gate and how the detuning $\vert \Delta_\text{R}^{ij} \vert$ is calculated. (a) A plot of an entire SLP, with both hyperfine levels as the blue line and the light blue line (left axis) and an enlargement of the hyperfine splitting as the dashed black line (right axis). The section marked by the thin black dashed line is enlarged in (b) and (c). The values used to plot are $A = 0.28$, $\lambda_1 = \SI{1064}{\nano \meter}$, $\lambda_2 = 4/5 \lambda_1 = \SI{851.2}{\nano \meter}$ and $\mathcal{P}_1 = \mathcal{P}_2 = 1$. (b) When exposed to precisely controlled microwave radiation the target atom is switched while keeping the switched population of neighbouring atoms under a threshold $P_\text{t}$. The positions of the atoms are calculated by taking the potential minima. (c) The positions of the atoms are used to calculate the hyperfine splitting of both atoms, and the detuning is the difference in this splitting, which in this example is $\hbar \vert \Delta_\text{R}^{ij} \vert / \eta E_\text{r} = 0.016$.}
	\label{fig:singleQubitFig}
\end{figure}

The potential in a SLP is calculated for both hyperfine levels, and the difference $- \vert \Delta U (x) \vert$ is plotted in Fig.~\ref{fig:singleQubitFig}(a). This difference is similar in form to the SLP itself. To calculate the site dependent detunings, the potential minima of all wells within a SLP are found, as in Fig.~\ref{fig:singleQubitFig}(b). For all atoms, $\Delta U_i$ can then be found and the detuning $\hbar \Delta^{ij}_\text{R} / \eta E_\text{r}$ is given as the difference in hyperfine splitting as seen in Fig.~\ref{fig:singleQubitFig}(c). The smallest of all $\Delta^{ij}_\text{R}$ sets the threshold, and thus only that detuning is considered. Note that the potential minima of the superlattice do not exactly match the corresponding minima of $- \vert \Delta U (x) \vert$, which tends to increase the detunings.
Additionally, the minima of the different hyperfine levels do not match either, although, for the red detunings considered here, this position shift is typically several orders of magnitude smaller than the laser wavelengths. Since there is no position shift for the deepest well any neighboring well shifts only serve to restrict the transition of non--target atoms even further.

\begin{figure}[tb]
		\includegraphics[height=105pt]{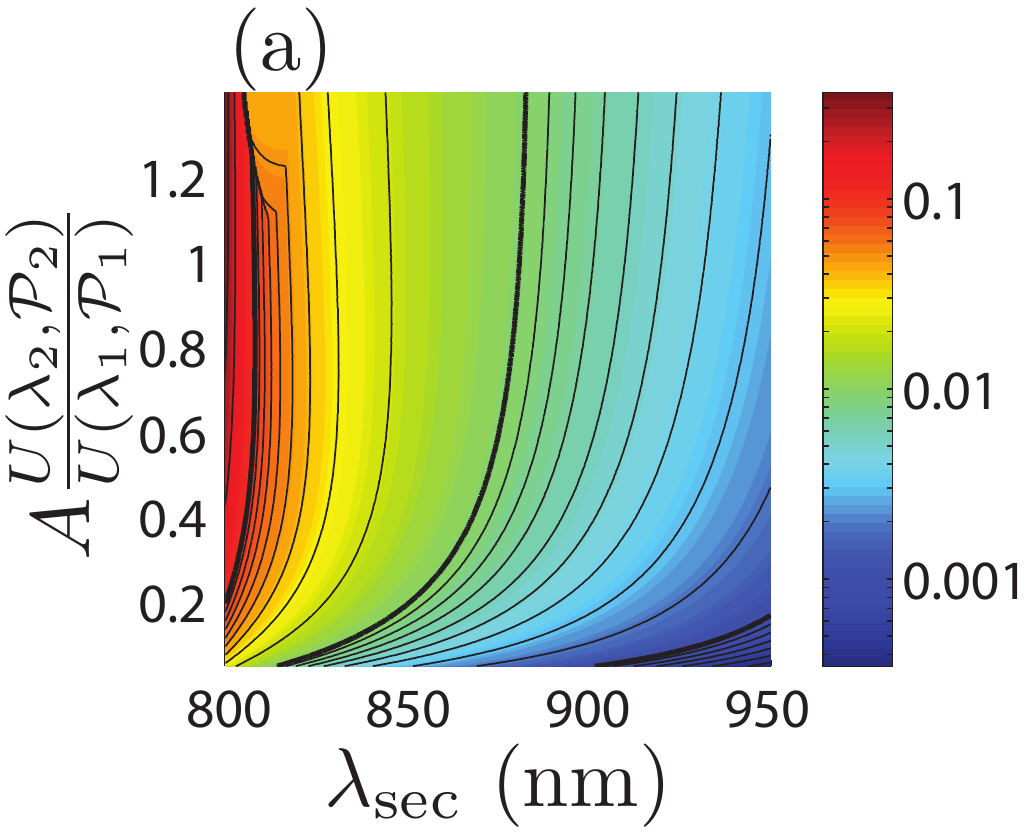}
		\includegraphics[height=105pt]{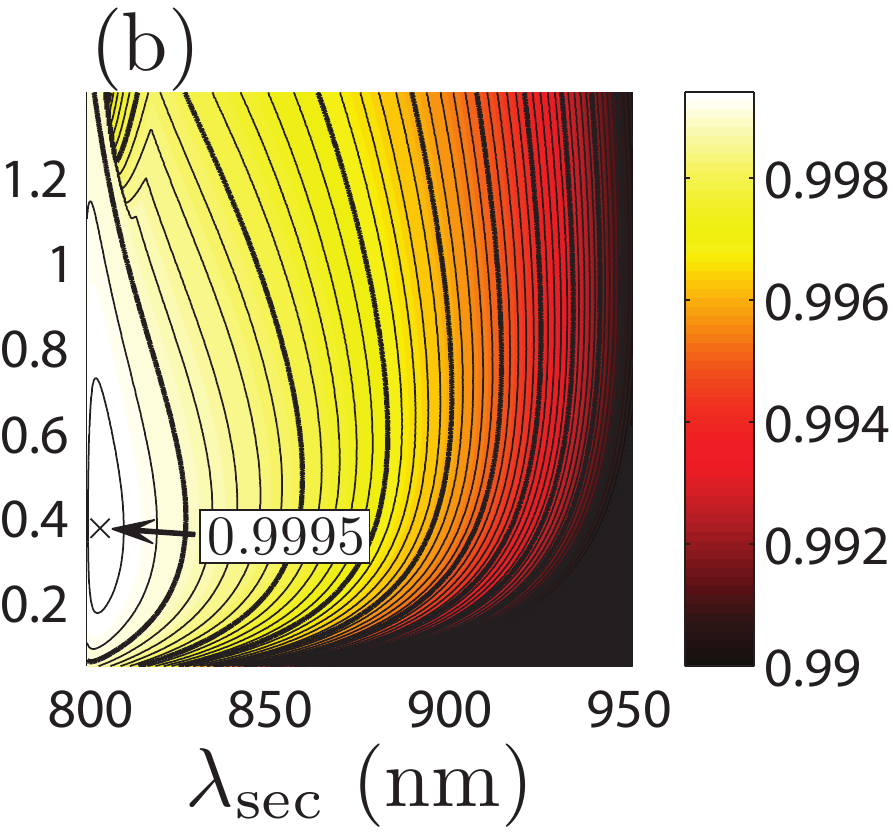}
	\caption{(Color online) Results for single-qubit gate calculations. (a)~The absolute detuning $\hbar \vert \Delta^{ij}_\text{R} \vert / \eta E_\text{r}$ calculated for an array of different wavelengths and secondary lattice depths. The contour lines represent logarithmic scaling. The addressing time can be calculated through the detuning via $t_\text{a} = \pi /\sqrt{P_\text{t}}\vert \Delta_\text{R}^{ij} \vert$ with $E_\text{r}/h \approx \SI{2}{\kilo \hertz}$. (b)~The corresponding probabilities of an operation without scattering the target atom. There is a maximum of high probability 0.9995 and the thick and thin contour lines represent steps of 0.001 and 0.0002. The black area represent probabilities beneath 0.99. Calculated with $\lambda_1 = \SI{1064}{\nano \meter}$ and $\mathcal{P}_1 = \mathcal{P}_2 = 1$.}
	\label{fig:singleQubitResults}
\end{figure}

The detuning for a range of secondary lattice wavelengths and lattice depths is seen in Fig.~\ref{fig:singleQubitResults}(a). Using $E_\text{r}/h \approx \SI{2}{\kilo \hertz}$ the largest shifts of Fig.~\ref{fig:singleQubitResults}~(a) result in gate time $\sim100\mathrm{ms}/\eta$. Since lattices of 100-1000$E_\text{r}$ can be realized routinely using high power lasers, gate times $0.1-1~\text{ms}$ should be feasible.
The largest detunings are seen close to the D$_1$ line and are generally larger when the lattice depths are similar.
This seems to suggest that the single qubit gate should be performed at the lowest possible detuning. This conclusion changes when the probability of scattering a photon $p_{sc} = \exp(-\gamma_{\text{sc}}t_\text{a})$, during a gate operation is included. The scattering rate is calculated using~\cite{Grimm}:
\begin{equation}
\gamma_{\text{sc}}(\textbf{r}) = \frac{\pi c^2 \Gamma^2}{2 \hbar \omega_0^3} \left( \frac{2}{\Delta_\text{2,F}^2} + \frac{1}{\Delta_\text{1,F}^2} \right) I(\textbf{r}),
\label{eq:scat}
\end{equation}
and rewriting the expression similarly to Eq. (\ref{eq:potential_num}) into $\gamma_{\text{sc}}/E_\text{r}$, including $A$ and $\eta$. In calculating the scattering rate, three fixed retro--reflected lasers with equal intensities and wavelengths, one for each dimension, are included plus the secondary laser in a single dimension. When calculating  $p_{sc}$ only the target atom is taken into consideration. As both the detuning $\Delta_\text{R}^{ij}$ and scattering rate scale linearly with lattice depth, the gate--time scattering probability is independent of the depth and the scaling factor $\eta$.

The probabilities of a successful operation $1-p_\text{sc}$ are mapped in Fig.~\ref{fig:singleQubitResults}(b) for the detunings calculated in Fig.~\ref{fig:singleQubitResults}(a).
At the optimum the maximum--probability of $1 - p_\text{sc} = 0.9995$ is reached. Increasing the primary wavelength increases the probability slightly, however, we have chosen to represent the results corresponding to $\lambda_1=1064\text{nm}$ due to the high availability of such a laser system.
Other polarizations have been tested, and the detunings $\Delta_\text{R}^{ij}$ were examined for $j$ not being the atom in the deepest well; both yielding similar or worse results than those presented above.
 When scaling up to a longer period SL a naive estimate of the total probability of not scattering an atom for $N$ atoms is the $N^{\textrm{th}}$ power of the probability of not scattering the target atom. The result would scale poorly with hundreds of atoms as $0.9995^{100} \sim 0.95$. As can be seen in Fig.~\ref{fig:singleQubitFig}(a), however, the intensity will decrease away from the maximum one resulting in a reduced error probability at larger distances.

\section{Two-qubit gate}\label{sec:two}
Having initialized an array of atoms in specific lattice sites, a two-qubit gate on a selected pair of neighboring atoms can be performed by exploiting the spin-exchange interaction. To achieve such a gate in this architecture requires the merging of two atoms in the same lattice site such that their wave functions overlap, as sketched in Fig.~\ref{fig:overview}(c).  The optimization of this non-trivial merger process is the subject of section ~\ref{section:merging}. In this section, we describe the gate mechanism and identify the requirements of performing such a gate in an optical superlattice. To minimize the gate time, we also numerically optimize the lattice depth and phase throughout the merging process and consider the detrimental effects of experimental preparation errors.

\subsection{Gate description}
The mechanism driving the two-qubit gate is the mutual interaction between two overlapping atoms which leads to spin exchange~\cite{Hayes2007,Weitenberg2011b,Anderlini2007}. Two initially separated qubits are combined in the same well in the ground and first excited vibrational levels of the well
\begin{equation}
\begin{split}
\alpha \Ket{1}_\text{L} + \beta \Ket{0}_\text{L} \rightarrow \alpha \Ket{1}_\text{g} + \beta \Ket{0}_\text{g}, \\
\tilde{\alpha} \Ket{1}_\text{R} + \tilde{\beta} \Ket{0}_\text{R} \rightarrow \tilde{\alpha} \Ket{1}_\text{e} + \tilde{\beta} \Ket{0}_\text{e},
\end{split}
\end{equation}
where $\Ket{1}$ and $\Ket{0}$ denote the spin based qubit states, $\alpha$, $\beta$, $\tilde{\alpha}$ and $\tilde{\beta}$ are the amplitudes, $\Ket{\cdot}_\text{L}$ and $\Ket{\cdot}_\text{R}$ denote the wave functions of the atoms in the left and right well, and $\Ket{\cdot}_\text{g}$ and $\Ket{\cdot}_\text{e}$ denote the wave functions of the atoms in the ground and excited vibrational levels of the merged well.

The two atoms in the merged well are identical bosons, so the two-particle wave function is symmetric under particle exchange. The new eigenenergy basis of the system is formed by the singlet and triplet states
\begin{equation}
\begin{split}
\Ket{s} & = \frac{1}{\sqrt{2}} \left( \Ket{1}_\text{g} \Ket{0}_\text{e} - \Ket{0}_\text{g} \Ket{1}_\text{e} \right), \\
\Ket{t_{-1}} & = \Ket{0}_\text{g} \Ket{0}_\text{e}, \\
\Ket{t_0} & = \frac{1}{\sqrt{2}} \left( \Ket{1}_\text{g} \Ket{0}_\text{e} + \Ket{0}_\text{g} \Ket{1}_\text{e} \right), \\
\Ket{t_{+1}} & = \Ket{1}_\text{g} \Ket{1}_\text{e}.
\end{split}
\end{equation}
The two-qubit state of the atoms can now be expressed via the basis of singlet/triplet states as $\Ket{1}_\text{g} \Ket{0}_\text{e} = (\Ket{t_0} + \Ket{s})/\sqrt{2}$ and $\Ket{0}_\text{g} \Ket{1}_\text{e} = (\Ket{t_0} - \Ket{s})/\sqrt{2}$.

The singlet spin state $\Ket{s}$ is antisymmetric, and hence its spatial wave function must be antisymmetric as well. In this wave function there is no density overlap between the two particles. The ultracold atoms primarily interact by contact, which means that there is negligible interaction in the state $\Ket{s}$. However, the wave function of the symmetric spin state $\Ket{t_0}$ must be symmetric, which leads to an interaction between the atoms and hence a change in energy $U_\text{int}$ when compared to the state $\Ket{s}$.

As the two-qubit state $\Psi$ evolve in time, the energy shift between the two states $\Ket{s}$ and $\Ket{t_0}$ will induce a phase shift
\begin{equation}
\Psi(t) = \frac{1}{\sqrt{2}} (e^{i U_\text{int} t / \hbar} \Ket{t_0} + \Ket{s}),
\end{equation}
which will induce periodic oscillations between $\Ket{1}_\text{g} \Ket{0}_\text{e}$ and $\Ket{0}_\text{g} \Ket{1}_\text{e}$. At time $T_\textsc{swap} = \pi \hbar / U_\text{int}$ the spin states are swapped and at time $T_{\sqrt{\textsc{swap}}} = \pi \hbar / 2 U_\text{int}$ the entangling $\sqrt{\textsc{swap}}$ gate is implemented, which is universal for quantum computation. The qubits can subsequently be separated by reversing the merging operation.

The gate time is set by the interaction between two $^{87}$Rb atoms and can be modelled by an effective 1D contact potential~\cite{Olshanii1998,DeChiara2008}
\begin{equation}
V_{\text{int}} \left( \vert x_1 - x_2 \vert \right) = g_{\text{1D}} \delta  \hspace{-2pt} \left( x_1 - x_2 \right),
\label{eq:Vint}
\end{equation}
where $x_1$ and $x_2$ are the coordinates of the two atoms, $\delta$ is the Dirac delta function and $g_{\text{1D}}$ is the effective 1D coupling strength. This strength is given by $g_{\text{1D}} = 2 a_{\text{s}} h \sqrt{\nu_y \nu_z}$ where $a_{\text{s}}$ is the scattering length of the atoms, $h$ is the Planck constant and $\nu_y$ and $\nu_z$ are the trap frequencies in the $y$-- and $z$--directions. For  $^{87}$Rb, the scattering length $a_{\text{s}}= 110 a_0$, where $a_0$ is the Bohr radius \cite{Boesten1997}.

\subsection{Lattice site merging}
\label{section:merging}
 \begin{figure}[tb]
		\includegraphics[width=\columnwidth]{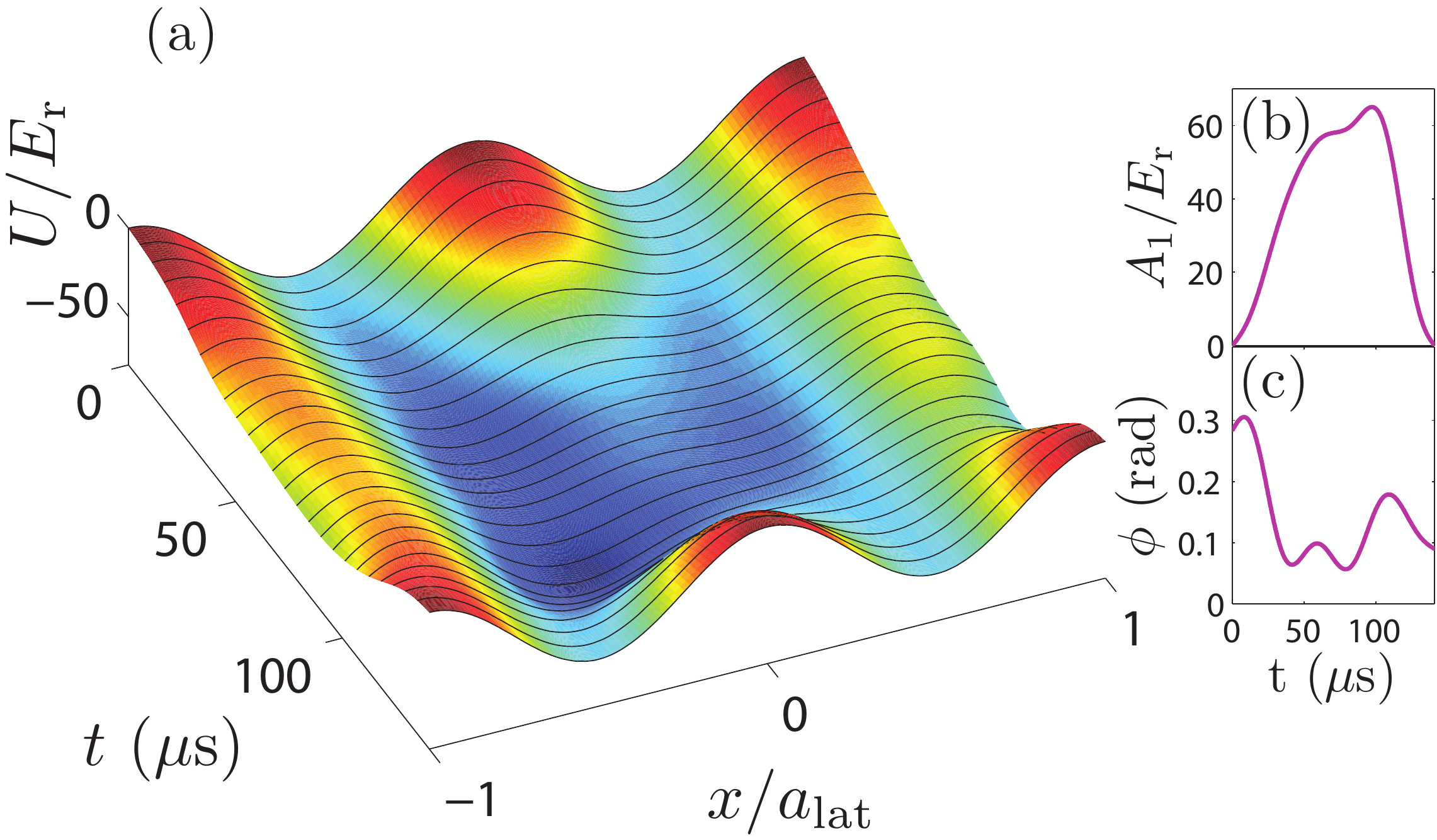}
	\\	\vspace{5pt}
		\includegraphics[width=.9\columnwidth]{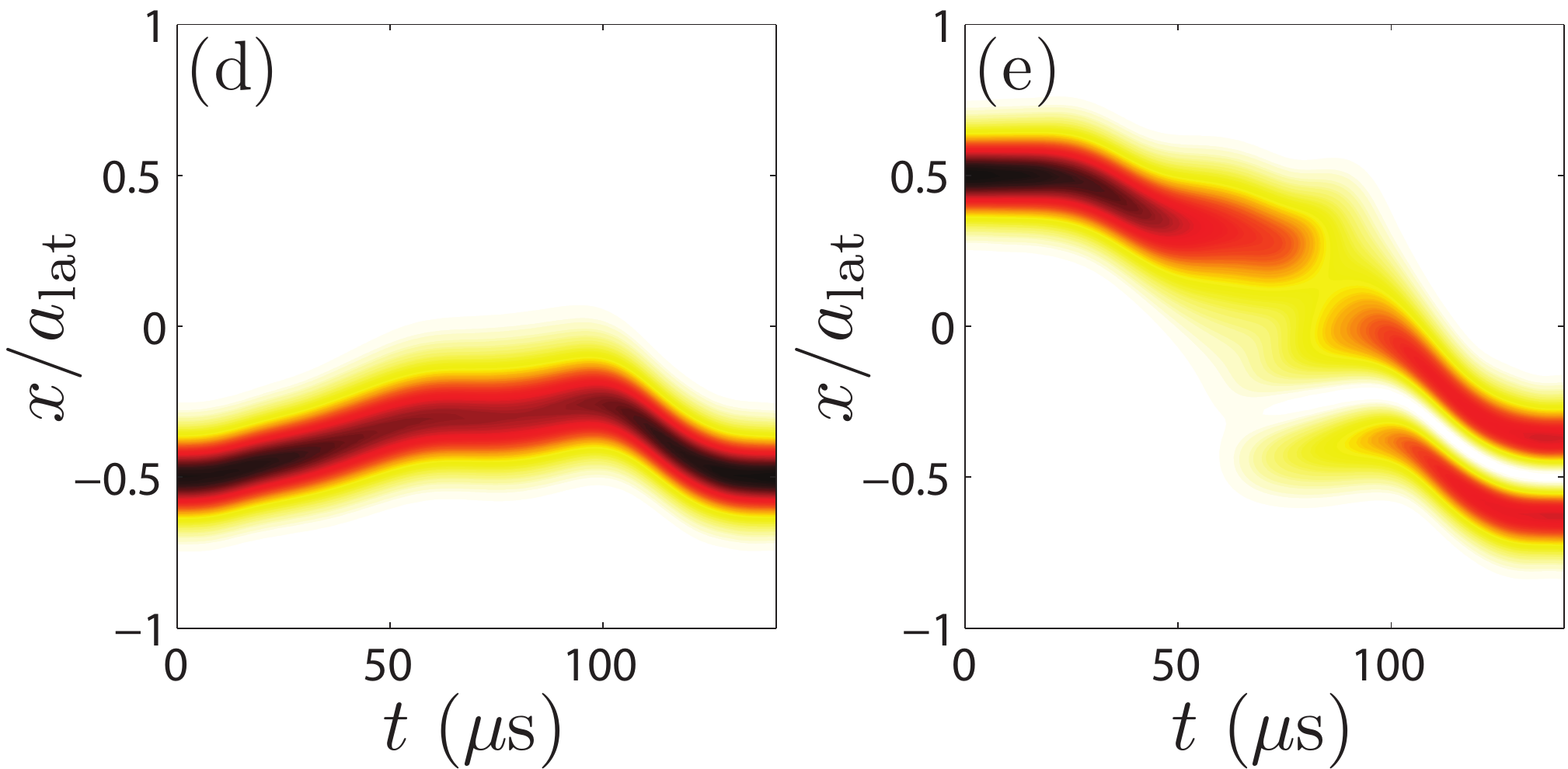}
	\caption{(Color online) An example of the operation required to realize a two-qubit gate. (a) The time--evolution of the potential leading to two atoms being sent in to one well. (b)-(c) Optimal control pulses for the amplitude and phase of the primary lattice leading to the potential deformation seen in (a). Density profiles of the two atoms as a function of time illustrating the mapping of one atom into the excited state of a neighboring well (e), while the other atom ends in the ground state (d).}
	\label{fig:twoQubitEx}
\end{figure}
In this section we will show how, by controlling the phase and depth of an optical superlattice, one can merge pairs of interacting atoms into a single lattice site in which they can perform the SWAP gate described above. An illustration of the superlattice potential during the merging process is seen in Fig.~\ref{fig:twoQubitEx}(a) with the corresponding values of amplitude and phase of the added lattice seen in (b) and (c). The potential minimum of the right well is shifted towards the $-x$ direction so that both atoms move into the well at $x=-0.5 a_\text{lat}$. This is shown in Fig.~\ref{fig:twoQubitEx}(d) and (e) where the density profiles of the two atoms are seen as a function of time. After being initially separate, one atom is promoted to the first excited vibrational state while the other remains in the ground state.
 \begin{figure*}[tb]
 		\includegraphics[width=.315\textwidth]{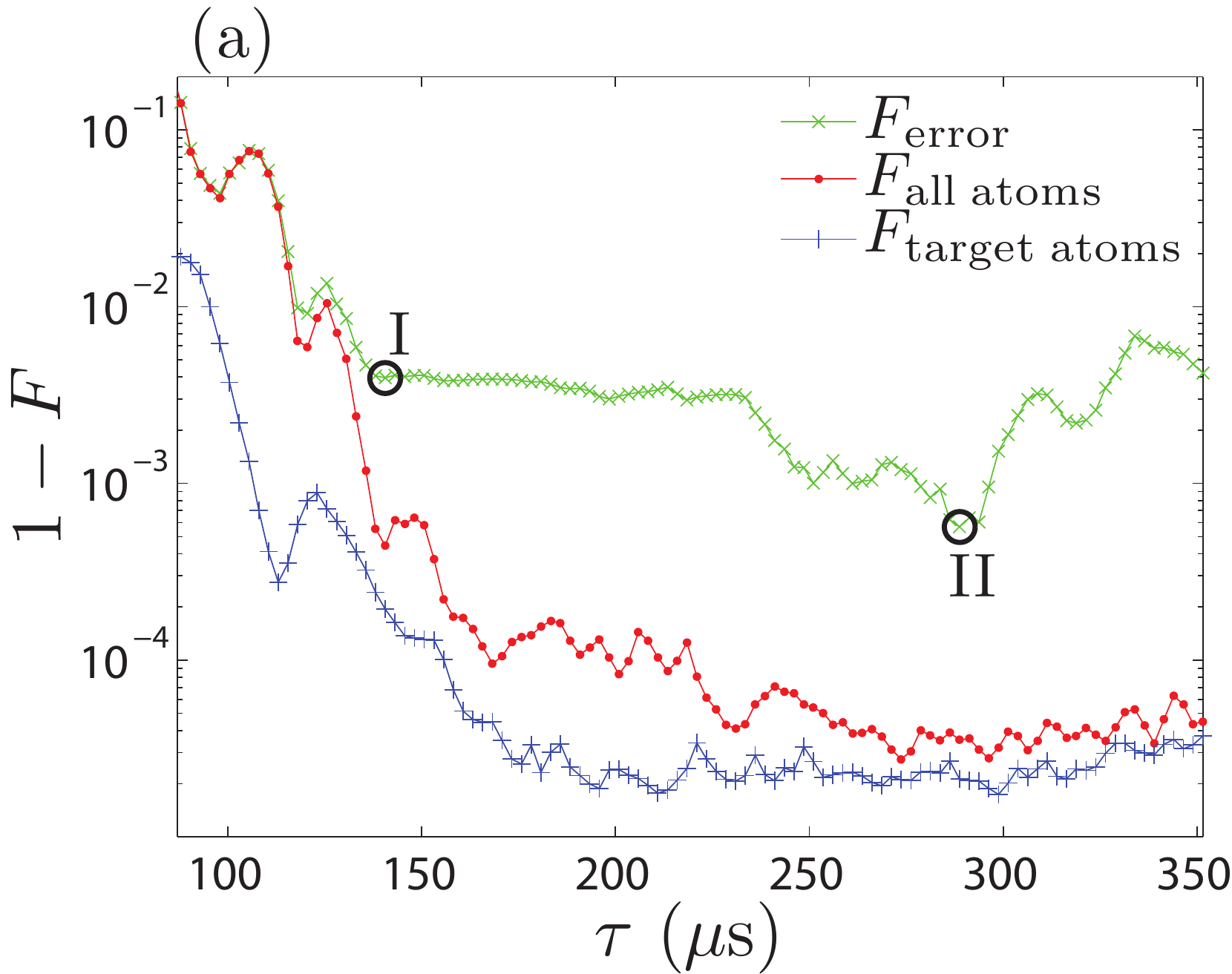} \hspace{12pt}
 		\includegraphics[width=.33\textwidth]{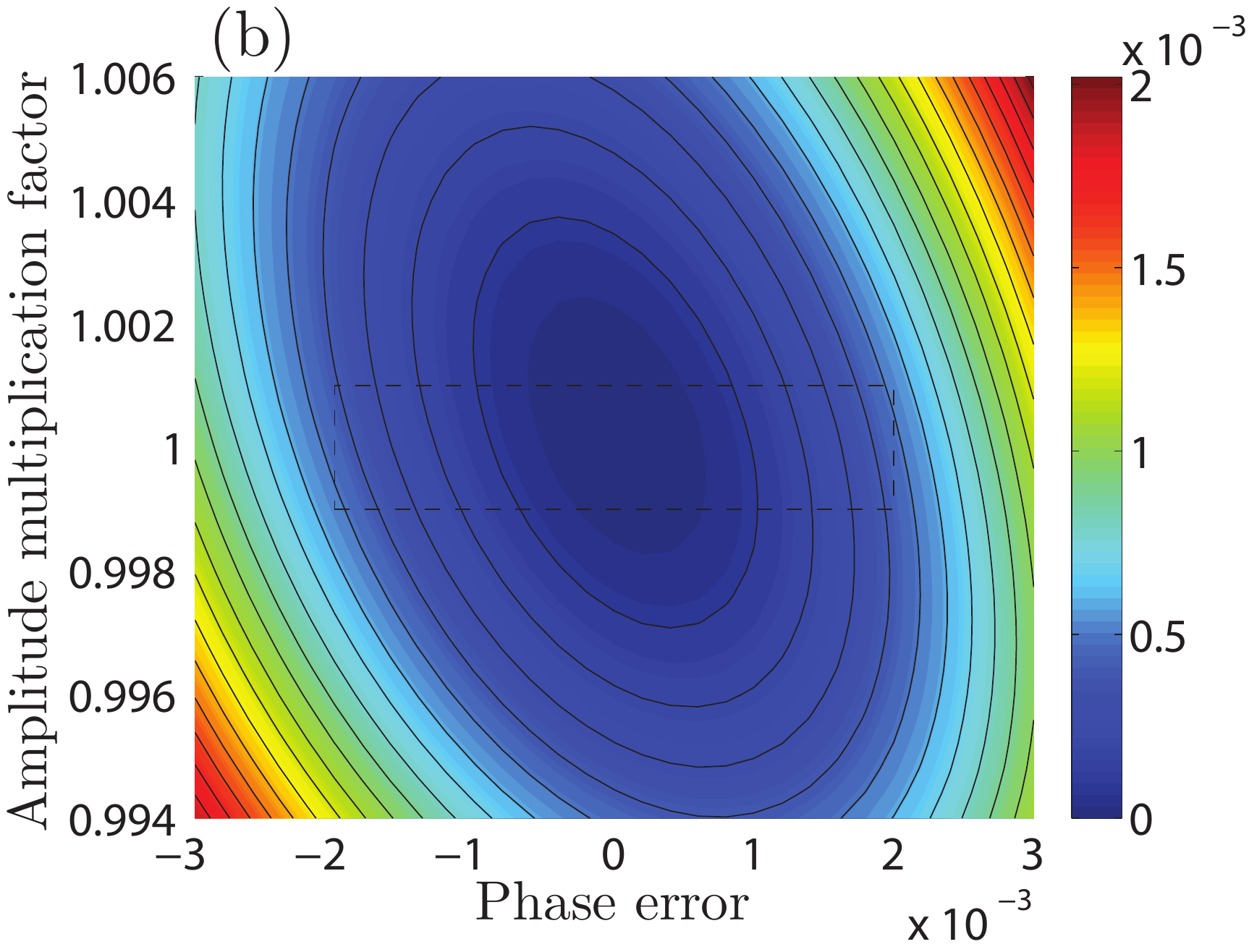} 
		\includegraphics[width=.315\textwidth]{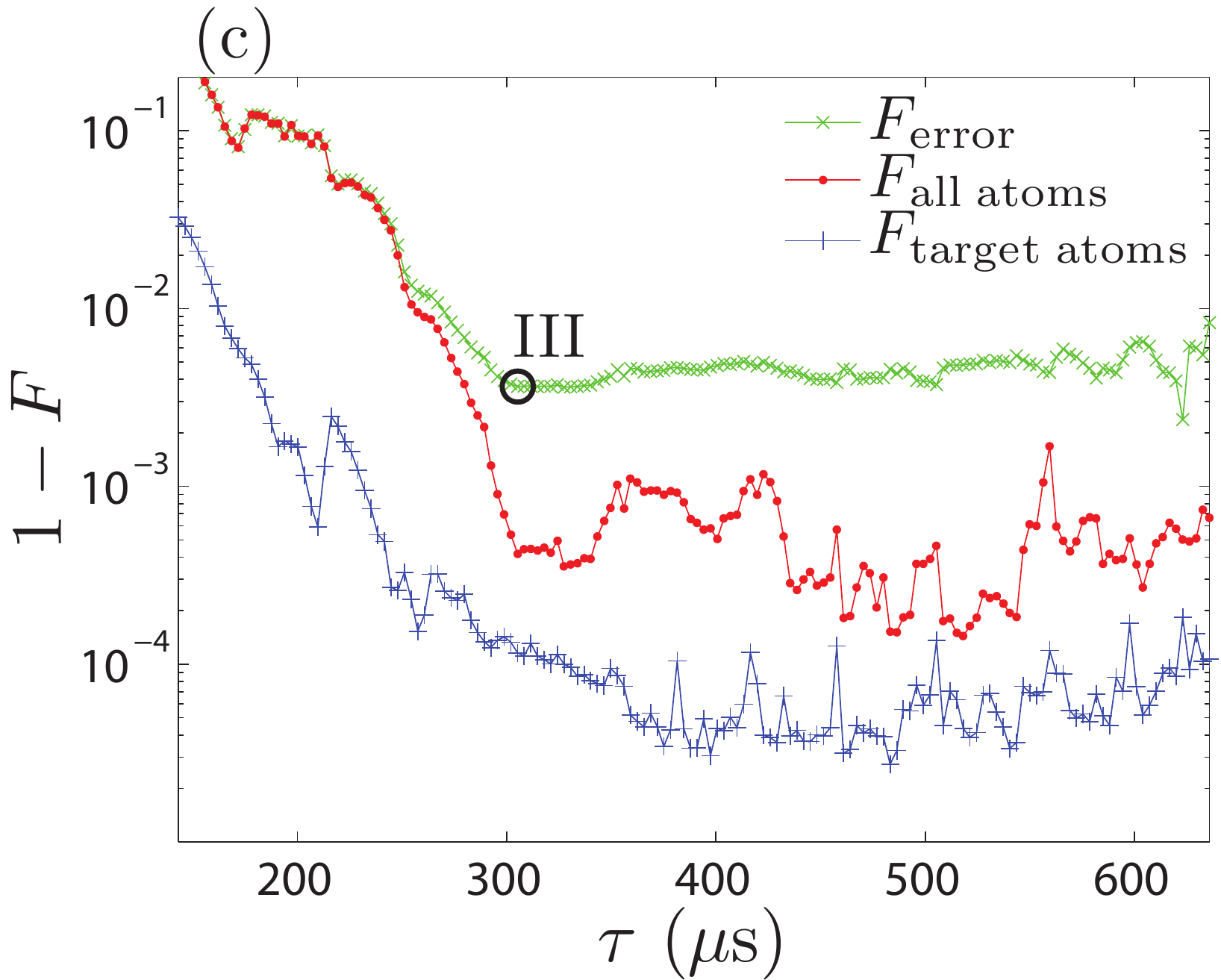}
	\caption{(Color online) Fidelity, $F$ for the simulations of the superlattice two-qubit gate. (a) Different fidelities obtained at different operation times with $n=5$, with two results of interest highlighted as I and II. The optimized merging scheme at II corresponds to the example shown in Fig.~\ref{fig:twoQubitEx}. (b) The change in fidelity for II as a function of error also illustrating how $F_\text{error}$ is calculated, i.e. by including all fidelities within the dashed lines and assuming the worst. The contour lines represent steps of $10^{-4}$. (c) Results for the lattice configuration $n=10$ with one more point of interest highlighted as III. The lattice beam wavelength, $\lambda_2$ is fixed at 1064~nm. Further results for I, II and III are seen in Table~\ref{tbl:twoQubitResults}.}
	\label{fig:twoQubitResults}
\end{figure*}
Starting from the total optical lattice potential:
\begin{equation}
\frac{U(x)}{E_\text{r} (\lambda_2)} = -A_1 \cos^2 \left( k_1x + \phi \right) - A_2 \cos^2 \left( k_2x + \pi/2 \right),
\end{equation}
we search for the optimum values of phase and depth that merges atoms with the highest fidelity. This is achieved by using the split-step method to simulate the time-evolution of the atoms and optimizing using a simplex algorithm. During an operation, the primary laser phase, $\phi{}$ and depth $A_1$ are varied while the secondary laser phase and depth are fixed at $\pi/2$ and 1~$E_\text{r}(\lambda_2)$. A primary phase of {\bf{$\phi = 0$}} merges the wells at $x \approx 0$. Note that when combining atoms into a single well,  merging is more easily achieved when a longer wavelength lattice is added to an initially populated short-wavelength lattice.

Precise experimental control of the lattice phases can be achieved in a retro-reflected optical lattice geometry by varying the primary beam frequency, $\Delta \nu$ to give a  phase change $\Delta \phi = 2 \pi d \Delta \nu / c$, where $d$ is the distance to the retro--reflector mirror. To achieve $\Delta \phi = \pi$ with $\lambda = \SI{1064}{\nano \meter}$ and $d=\SI{1}{\meter}$, a change in frequency $\Delta \nu  = \SI{150}{\mega \hertz}$ is required. The fact that the dynamics is controlled using the laser frequency - one of the most well controlled quantities in physics - illustrates one of the appealing features of our proposal.

We choose to independently optimize the merger for three different fidelity classifications. The first, $F_\text{target atoms}=P_gP_e$, is the population of the two target atoms in the ground and excited states. To reflect the effects of the merger sequence on non-target atoms, we optimize a second fidelity, $F_\text{all atoms}=P_gP_e\prod_i{}P_{g,i}$ where $i$ are the atoms in each SLP which are not involved in the gate. For each simulation, experimental sources of errors are added to the time dependent amplitude and phase of the added lattice corresponding to error in intensity and frequency. The sources of noise are assumed to be of a sufficiently low frequency to be considered constant during the operation and are therefore incorporated by adding a global shift to the obtained control pulses.
Based on this, a third fidelity, $F_\text{error}$, is optimized which takes the worst obtained fidelity within the array of errors used, also including all atoms in the SLP.

The existence of several local maxima in the optimizational landscape neccessitated optimization starting from long times moving towards shorter times and vice-versa. At each point in time, the highest fidelity was selected.

Two secondary lattices wavelengths $\lambda_2 = \SI{851.2}{\nano \meter}$ and $\SI{957.6}{\nano \meter}$ and a primary lattice wavelength of $\lambda_1 = \SI{1064}{\nano \meter}$ are studied. These wavelengths correspond to $n=5$ and $n=10$ superlattices. The error boundaries used to optimize  $F_\text{error}$ are set to 0.1\% for amplitude and a phase offset of 0.2\%, as shown by the box in Fig. ~\ref{fig:twoQubitResults}(b). As can be seen, with appropriate control of the phase (i.e. the relative frequency difference) one can tolerate power fluctuations of the order of 1\%, while still remaining below $10^{-3}$ infidelity.

The total gate times for the \textsc{swap} and $\sqrt{\textsc{swap}}$ gates are set by calculating the interaction during the merging operation. For the \textsc{swap} ($\sqrt{\textsc{swap}}$) gate, the interaction induced phase shift is required to be $n_i\pi$ ($\frac{n_i\pi}{2}$), where $n_i$ is an integer. The total phase shift picked up during the merging operation will also be picked up when reversing the merging operation to split up the atoms into separate wells again. When requiring a certain phase shift, a total gate time is then given by twice the operation time $\tau$ plus a time given by the stationary interaction. From the total gate times, the probabilities of scattering an atom in the SLP during a gate is calculated including spatially varying intensity and $\SI{1064}{\nano \meter}$ lattice in the $y$- and $z$-dimension with a depth of $32E_\text{r}$.

The resulting fidelities at different operation times $\tau$ are seen in Fig.~\ref{fig:twoQubitResults}(a) and (c). Three points of interest are I, II and III, and total gate times and scattering rates for these operations are shown in Table~\ref{tbl:twoQubitResults}.

\begin{table}[tb]
\centering
\caption{Further results for the three highlighted merging processes I, II and III from Fig.~\ref{fig:twoQubitResults}. The total gate times $T_\textsc{SWAP}$ and $T_{\sqrt{\textsc{SWAP}}}$ are calculated by requiring the total interaction to cause a fixed phase shift. From the the merging process, the probabilities $P_\text{sc, \textsc{swap}}$ and $P_{\text{sc, } \sqrt{\textsc{swap}}}$ of scattering an atom within the SLP during a gate are also calculated.}
\begin{tabular}{l | c c c c c c}
\toprule
& $\tau (\SI{}{\micro \second})$ & $F_\text{error}$ & $T_{\sqrt{\textsc{swap}}} (\SI{}{\micro \second})$  & $T_\textsc{swap} (\SI{}{\micro \second})$ & $P_{\text{sc, } \sqrt{\textsc{swap}}}$ & $P_\text{sc, \textsc{swap}}$  \\
\hline
I & 141 & $0.9960$ & $460$ & $366$ & $4.0 \times 10^{-4}$ & $3.2 \times 10^{-4}$ \\
II & 289 & $0.9994$ & $634$ & $728$ & $5.5 \times 10^{-4}$ & $6.3 \times 10^{-4}$ \\
III & 305 & $0.9964$ & $769$ & $638$ & $8.3 \times 10^{-4}$ & $7.0 \times 10^{-4}$ \\
\hline
\end{tabular}
\label{tbl:twoQubitResults}
\end{table}

Merging sequence II is also depicted in Fig.~\ref{fig:twoQubitEx}. In this case the operation is plotted without errors included. The change in fidelity when including various errors for II is seen in Fig.~\ref{fig:twoQubitResults}(b) where the area marked by the dashed line represent the errors included to optimize $F_\text{error}$. This high fidelity result is achieved at a modest operation time of less than 300 $\mu{}$s. Even choosing a larger SLP corresponding to $n=10$, the operation time is comparable at a slightly reduced fidelity. This illustrates that larger qubit registers are feasible.

\section{Conclusion}
\label{sec:conc}

We have presented a novel architecture for quantum computing using the spatially dependent potential of neutral atoms in long periodicity optical superlattices implemented by superposing two optical lattices with close-lying periodicity. We have identified the fastest possible single-qubit gate times given a maximum tolerable rotation error on the remaining atoms at various different values of the lattice wavelengths. Including the detrimental effect of spontaneous emission, we show that gates in the sub-millisecond regime can be realized with less than $10^{-3}$ total error probability. The proposed two-qubit gate takes advantage of the fact that at the node of the superlattice period there is an isolated double well system in which merger can be realized by controlling the relative intensity and frequency of the two lattices. Controlling the relative phase of the two lattices the node can be positioned at an arbitrary pair of wells. We numerically optimize the merger to implement an entangling $\sqrt{\textsc{swap}}$ two-qubit gate. Including realistic sources of error and the accumulated errors of atoms not participating in the merger we still obtain total gate error probabilities of the order of $10^{-3}$ with periodicities up to $n=10$.
Future work will focus on extending the merging scheme to fractional $n$ superlattices to achieve selectivity across even larger qubit registers and the optimization of custom pulse protocols~\cite{Lee2013} to increase the single qubit gate robustness. Finally, we would like to point out that although this work has focussed on the manipulation of individual atoms the method could also be used to select a single plane in a one dimensional lattices as an alternative to current techniques relying on magnetic field addressing~\cite{Schrader2004,Sherson2010}.

\section*{Acknowledgments}
The authors acknowledge support from the Danish Council for Independent Research, Natural Sciences, the Lundbeck Foundation and a Marie Curie IEF in FP7.
\label{sec:ack}

\bibliography{references2}

\end{document}